# GRAVITATIONAL LENSING IN CLUSTERS OF GALAXIES: NEW CLUES REGARDING THE DYNAMICS OF INTRACLUSTER GAS

Jordi Miralda-Escudé[1,2] & Arif Babul[3]

[1]Institute for Advanced Study, Princeton, NJ 08540, USA

[2]Institute of Astronomy, Madingley Road, Cambridge CB3 0HA, UK

[3]Canadian Institute for Theoretical Astrophysics, 60 St George St., Toronto, CANADA. M5S 1A1

E-mail: jordi@guinness.ias.edu, babul@cita.utoronto.ca

## ABSTRACT

Long arcs in clusters of galaxies, produced by gravitational lensing, can be used to estimate the mass interior to the arcs and therefore, constrain the cluster mass distribution. The radial density distribution of the intracluster gas can be extracted from the X-ray surface brightness observations. If the gas temperature is also known, it is then possible to probe the dynamical state of the gas and in particular, to test the hypothesis that the intracluster gas is in hydrostatic equilibrium within the gravitational potential of the cluster as a result of thermal pressure support.

We analyze three clusters that exhibit large arcs, whose X-ray surface brightness profiles have been observed, and whose gas temperatures have been determined. In two of the clusters, A2218 and A1689, the central mass implied by lensing is a factor of 2–2.5 too large for the gas at the observed temperature to be in hydrostatic equilibrium solely due to thermal pressure support. In other words, if we accept the mass estimate derived from the lensing analysis and demand that the X-ray surface brightness profile be consistent with the observations, the temperature of the gas has to be a factor of 2–2.5 larger than the observed value. The results for the third cluster, A2163 (the most luminous and the hottest cluster known), are more ambiguous.

The discrepancy between the X-ray and the lensing mass estimates arise because the presence of arcs imply a highly concentrated cluster mass distribution whereas the observed X-ray profiles imply a more extended mass distribution. The large X-ray core radii are not the result of the limited resolution of the X-ray detectors. We consider various possibilities that could account for the discrepancy.

*Subject headings*: gravitational lenses - galaxies: clustering - X-rays

## 1. INTRODUCTION

It has long been recognized that mass concentrations can gravitationally lens light rays from background sources and that such images offer the possibility of probing the matter distribution in the lensing bodies (*e.g.* Einstein 1936; Zwicky 1937). Soon after the discovery of the first gravitational lenses (Walsh *et al.* 1979), several authors considered the possibility that clusters of galaxies could act as gravitational lenses, noting that if the cluster mass distribution is sufficiently concentrated towards the center, they would be able to produce images with splitting angles as large as 60" (Turner, Ostriker, & Gott 1984; Narayan, Blandford, & Nityananda 1984). The discovery of arcs in clusters of galaxies, and their subsequent confirmation as gravitationally lensed images of background galaxies (Soucail *et al.* 1987; Lynds & Petrosian 1989), provided the first hints that the mass distribution in these clusters is indeed highly concentrated. Grossman and Narayan (1988) argued that the core radius for the mass distribution in lensing clusters had to be smaller than $\sim 100 h_{50}^{-1}$ kpc if the cluster velocity dispersion is assumed to be the same as that of the cluster galaxies. X-ray observations of the clusters, however, typically imply much larger core radii of $\sim 400 h_{50}^{-1}$ kpc (*e.g.* Jones & Forman 1984; Edge & Stewart 1991). Various explanations have been put forth to explain the different core sizes, including the possibility that the large X-ray cores are artifacts of the limited resolution of the X-ray detectors, and that both the gravitational lensing as well as X-ray observations may affected by the presence of unresolved substructure in the cluster cores (Grossman & Narayan 1989; Miralda 1993).

In this paper, we use the observed arcs to estimate the total mass within a central region in the clusters and in combination with the X-ray temperature and surface brightness measurements, attempt to probe the dynamical state of the hot intracluster gas. This allows us to test the generally accepted hypothesis that the hot gas in clusters is in hydrostatic equilibrium as a result of thermal pressure support. Specifically, we consider the three clusters: A2218, A1689, and A2163. Highly distorted arcs have been observed in all three. Furthermore, the X-ray properties, including the X-ray surface brightness profile and the temperature of the X-ray emitting intracluster gas, of the all three clusters have been measured and well-studied. The X-ray and the optical properties of the three clusters are summarized in Table 1. We use the observed arcs to estimate the total projected mass within the radius of the arcs, and therefore constrain the cluster potential. Since the temperature of the gas is also known, we then explore whether it is possible for the intracluster gas to exist in thermal pressure-supported hydrostatic equilibrium in the the constrained cluster potential, and produce X-ray surface brightness profiles consistent with those observed. Throughout this paper, we assume an Einstein-de Sitter model for the universe ($\Omega = 1$) and adopt a Hubble constant of $H_o = 50 \, \mathrm{km \, s^{-1} \, Mpc^{-1}}$. Our method will be described in § 2, and the results are presented in § 3. Possible interpretations of the results are discussed in § 4. Finally, in § 5 we summarize our findings and suggest directions for future theoretical and observational work.

## 2. METHOD OF ANALYSIS

In gravitational lensing, highly elongated arcs lie very close to the "critical line" of a cluster. In a spherical potential, the critical line traces a circle. Strictly speaking, a highly elongated arc in a spherically symmetric potential should always be accompanied by a counterarc of approximately the same length on the opposite side of the lens; however, the slightest deviation from spherical symmetry will cause the counterarc to shrink in size or even disappear altogether. Therefore, we will not be concerned about the counterarc and assume that the observed arc is coincident with the critical line.

The mean projected surface mass density within the critical circle of angular radius $b$ is

$$\bar{\Sigma}(b) = \Sigma_{crit} = \frac{c^2}{4\pi G} \left(\frac{D_s}{D_l D_{ls}}\right) , \qquad (1)$$

where $D_l$, $D_s$, and $D_{ls}$ are the angular diameter distances to the lens, to the source, and from the lens to the source, respectively (*e.g.* Blandford & Kochanek 1988). The projected mass within the critical circle (i.e., the mass contained within a cylinder of radius $b$ passing through the centre of the cluster) is

$$M_p(b) = \Sigma_{crit} \, \pi \, b^2 \, D_l^2 . \qquad (2)$$

The above relationship, between the radius of an arc (critical line) and the projected mass contained within, is relatively insensitive to small deviations from sphericity.

We consider two different parameterized density profiles for the mass distribution in the clusters:

$$\rho(r) = \frac{\rho_0}{[1 + (r/r_c)^2]^{\gamma/2}} \times \left(1 + F_{cD}\frac{r_c^2}{r^2 + r_{cD}^2}\right) . \qquad (3)$$

$$\rho(r) = \frac{\rho_0}{(r/r_c)^\alpha \, (1 + r/r_c)^{\gamma-\alpha}} . \qquad (4)$$

In both cases, the overall normalization, denoted by $\rho_0$, is fixed by equation (2) once all the other free parameters are defined. In the first profile (equation 3), the density falls off as $r^{-\gamma}$ at large radii, flattens at intermediate radii, and may steepen again at some smaller radii, simulating the influence of a cD galaxy whose importance is quantified by the parameter $F_{cD}$. In total, there are three free parameters: the core radius $r_c$, the slope at large radius $\gamma$, and $F_{cD}$. We fix the value of the galaxy core radius at $r_{cD} = 0\rlap{.}''1$; small deviations about this value have no effect on the observable properties (the parameter $r_{cD}$ is introduced only for ease in the numerical calculations). In the second profile, the density drops as $r^{-\alpha}$ for $r \ll r_c$ and as $r^{-\gamma}$ for $r \gg r_c$, where $r_c$ is the core radius. This mass profile also has three free parameters: $r_c$, $\gamma$, and $\alpha$.

We assume that the gas in the cluster potential defined by the above mass distributions is in hydrostatic equilibrium and obeys the polytropic equation of state: $p \propto \rho_g^{\gamma_T}$. The radial distribution of the gas is, therefore, prescribed by

$$\frac{d\log \rho_g}{d\log r} = -\frac{G\,M(r)\,\mu}{\gamma_T k T(r)\,r} \ , \qquad (5)$$

where $\mu$ is the mean mass per particle in the gas (we adopt a value of $\mu = 0.6 m_p$), and $M(r)$ is the total mass contained within a sphere of radius $r$. In most of our models, we will assume that the gas is isothermal (*i.e.* $\gamma_T = 1$). Although radial variations in the temperature are hard to detect, the few clusters whose temperature profiles have been measured indicate that the gas temperature is either constant over the region of interest to us, or tends to decrease towards the centre (Schwarz *et al.* 1992).

Once the gas density is determined, we compute the model cluster's X-ray surface brightness by approximating the thermal bremsstrahlung emissivity of the intracluster gas, in a fixed energy band, as $\epsilon \propto \rho_g^2\, T^{-0.3}$ and by integrating along the line of sight. In order to compare the computed X-ray surface brightness with the observations, we process the model results by convolving it with the detector point spread function, integrating the flux in radial bins, and adding noise. (We only consider noise in the counts of the cluster surface brightness. Our simulated noise for the points at large radius is therefore underestimated, but this does not affect our conclusions, which depend only on the concentration of the X-ray surface brightness at small radii.) We then vary the free parameters in the assumed mass profile until we achieve the best possible fit to the observations. We note that this procedure is completely equivalent to the way in which X-ray profiles have been analyzed in the literature but with the crucial difference that we enforce the mass constraint imposed by the lensing analysis (equation 2).

## 3. RESULTS

We begin with an analysis of A2218. A2218 is an Abell richness class 4 cluster at a redshift of $z_l = 0.175$ (LeBorgne *et al.* 1992), with an X-ray luminosity in the 2–10 keV band of $L_x = 1 \times 10^{45}$ erg s$^{-1}$ (David *et al.* 1993). The radial velocity dispersion of the cluster galaxies is 1370 km s$^{-1}$ (LeBorgne *et al.* 1992) and the cluster's X-ray temperature has been determined by McHardy *et al.* (1990) to be 6.7 keV or $7.8 \times 10^7$ K. For the present purposes, we will adopt a value of $T_x = 8 \times 10^7$ K. The optical image of A2218 (c.f. Pelló et al. 1988, 1992) reveals the presence of two seemingly distinct galaxy concentrations, the larger and the more luminous of the two swarms occurring around the brightest cluster galaxy, a cD galaxy. The central galaxy of the secondary clump is also very bright and is displaced by 67″ from the cD. The optical image also reveals the existence of several arcs around the cD galaxy as well as two arcs close to a central galaxy in the second clump. The location and the morphology of the latter two arcs suggest that the second mass concentration is much smaller than that around the cD galaxy. Furthermore, the X-ray peak in the ROSAT PSPC image of A2218 (Stewart et al. 1993) appears to be coincident with the location of the cD galaxy. Consequently, we designate the cD galaxy as the centre of the cluster and focus on the associated large arcs.

One of these arcs (object number 359 in Pelló et al. 1992) has a measured redshift of 0.702, and is 20″.8 from the centre. There is another object, number 328 in Le Borgne et al. (1992), which is very likely another image of the source producing the 359 arc, since it has the same colour and is close to the position where a second image ought to be in a cluster potential with the same ellipticity as the central galaxy (Kneib et al. 1993b). The ellipticity of this object 328 also agrees with the expected shear induced by the cluster at $z_s = 0.702$. If that is correct, then the arc 359 should be a double image on a fold catastrophe, with possibly a part of the arc being eclipsed by the cluster galaxy 373 (see Fig. 4 in Pelló et al. 1992). The critical line should, therefore, intersect this arc and we adopt $b = 20″.8$ as the critical radius in A2218 for a source at $z_s = 0.702$.

In Figure 1a, we juxtapose the observed X-ray surface profile of A2218 derived from the ROSAT PSPC image (Stewart et al. 1993) and the surface brightness profile corresponding to the mass distribution in Model 1 (see equation 3) that comes closest to reproducing the observations. In figure 1b, we do the same for Model 2 mass distribution (see equation 4). The open squares denote the observed surface brightness. The solid line is the surface brightness profile for the model, and the crosses show the result of convolving the model with the ROSAT PSPC point spread function (which was approximated as a gaussian with FWHM of 32"), integrating the flux in radial bins, and adding noise. The best-fit parameters for Model 1 are $\gamma = 3.3$, $r_c = 14″$, $F_{cD} = 0$. The best-fit parameters for Model 2 are $\gamma = 5.3$, $\alpha = 0$, and $r_c = 31″$. Clearly, the observed surface brightness profile is far too flat and extended to be compatible with the distribution of gas in pressure equilibrium at the observed temperature in a cluster potential where the amount of projected mass within the radius of the arc is given by equation (2). Furthermore, in neither of the models is the steepening of the mass density profile at small radii — to represent the influence of cD galaxy — preferred (i.e. $F_{cD} = 0$ in Model 1 and $\alpha = 0$ in Model 2). The addition of a central mass concentration only exacerbates the discrepancy between the observed and the model surface brightness profiles by causing the latter profile to steepen even more.

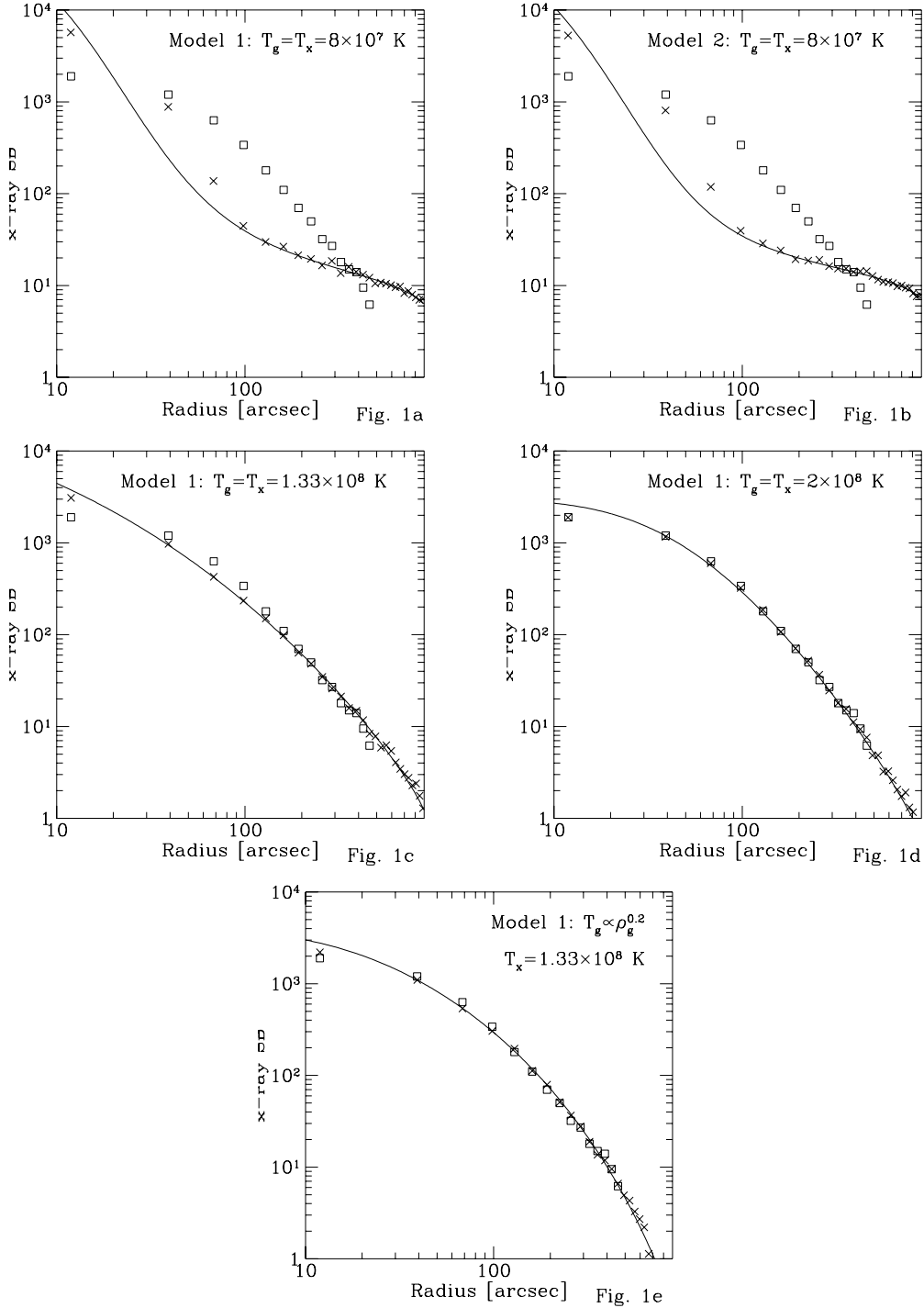

**Figure 1** — Model fits to the X-ray surface brightness profile (in counts cm$^{-2}$ s$^{-1}$) derived from the ROSAT PSPC image of A2218. The open squares are observations, the solid lines are the model results, and the crosses are the latter convolved with detector point spread function, integrated into bins, and with noise added. In (1a), we show the best fit obtained assuming an isothermal gas distribution, fixing the gas temperature to the observed X-ray temperature, and using Model 1 to parametrize the mass profile in the cluster. (1b) is the same, but for Model 2 mass profile. In (1c), we show results for Model 1 mass profile but with gas temperature a factor 1.6 higher than in (1a); a better fit is acquired because the higher temperature allows the gas to be more extended The fit in (1d) is even better; the gas temperature is a factor 2.5 higher. (1e) shows the result for polytropic gas ($T \propto \rho_g^{0.2}$) in the cluster potential defined by Model 1 mass. The emission-weighted spectral temperature is same as that for isothermal gas in (1c) though the fit is better. However, the temperature gradient implied by the model ought to have been observed.

The nature of the above discrepancy can be more easily understood if we assume that the cluster mass distribution is described by a singular isothermal sphere, $\rho \propto r^{-2}$, and has a velocity dispersion $\sigma$. The critical radius of such a cluster is (*c.f.* Blandford & Kochanek 1988)

$$b = \frac{4\pi\sigma^2}{c^2} \frac{D_{ls}}{D_s} \ . \qquad (6)$$

If the mass is assumed to have the same velocity dispersion as the gas, then $\sigma = (kT/\mu)^{1/2} = 1050\,\text{km}\,\text{s}^{-1}$, and the critical radius for a source at redshift $z_s = 0.7$ is $b = 21''$, roughly the same as that implied by the arcs. In this model, however, the gas profile is the same as that for the mass and the resulting X-ray profile diverges at small radii as $r^{-3}$. In order to flatten the X-ray profile within the central 30", a core of approximately the same size must be introduced in the mass distribution. However, if we require that the projected mass within 20" be unchanged (the lensing constraint), then the velocity dispersion of the cluster will be higher than that computed above. But since the temperature of the gas is fixed, the gas density profile will need to be steeper and more peaked in order to satisfy the conditions for hydrostatic equilibrium. Consequently, it is impossible to find a model with an isothermal gas distribution as extended as that implied by the observed X-ray surface brightness data, when both the gas temperature and the mass profile are constrained by X-ray and lensing observations.

Figure 1c shows the X-ray surface brightness profile for Model 1 mass distribution, with $F_{cD} = 0$, that best matches the observations when the gas temperature is raised by a factor of 1.66 above that observed (*i.e.* $T = 1.33 \times 10^8$ K). The mass distribution is characterized by $\gamma = 1.82$ and $r_c = 0''$. In comparison with the theoretical profile shown in Figure 1a, the current profile presents a much better fit to the observations. However, there is still some discrepancy in the sense that the observed X-ray profile is flatter and more extended in the central region than the predicted one. Raising the gas temperature by a factor of 2.5 above that observed (*i.e.* $T = 2 \times 10^8$ K) yields an acceptable fit (see Figure 1d; model parameters: $\gamma = 1.82$ and $r_c = 17''$). Consequently, the effective pressure of the intracluster has to be a factor of $\sim 2-2.5$ greater than the thermal pressure estimated on the basis of the X-ray temperature of the gas, in order to simultaneously fit both the observed X-ray surface brightness data and the gravitational lensing constraint. Alternatively, good fits to the observed X-ray surface brightness profile can be achieved by fixing the gas temperature at its observed value but reducing the projected mass internal to the critical radius by a factor of $\sim 2-2.5$ below that required for the formation of an elongated arc.

Thus far, we have only considered cases where the intracluster gas is isothermal. Figure 1e shows the X-ray surface brightness profile for intracluster gas obeying the polytropic equation of state (*i.e.* $p \propto \rho_g^{\gamma_T}$) and in hydrostatic equilibrium within a cluster potential defined by Model 1 mass distribution characterized by $F_{cD} = 0$, $\gamma = 2.0$ and $r_c = 7''$. The polytropic index of the gas distribution is $\gamma_T = 1.2$. The X-ray temperature of the gas in our model is same as for the model shown in Figure 1c ($T_x = 1.33 \times 10^8$ K, a factor of 1.66 higher than the observed value); however, the fit to the observed X-ray surface brightness profile is better. This is due to the fact the polytropic model has a large X-ray core, which arises because (1) the temperature rise towards the centre, in combination with the requirements of hydrostatic equilibrium, implies a relatively flat gas distribution, and (2) the X-ray emissivity depends on the gas temperature in such a way that for a gas at given density, a higher temperature leads to lower X-ray emissions. The high temperature of the gas in the central regions is not reflected by the X-ray temperature measurement since the latter is an emission-weighted temperature, averaged along each line of sight in the cluster and over all lines of sight. In contrast, the X-ray temperature of an isothermal gas is a direct measure of its actual temperature and a high gas temperature is required in order to have a gas distribution that is sufficiently extended so as to account for the large core in the observed X-ray surface brightness profile.

The presence of a temperature gradient in the polytrope model, however, is problematic. The decline in the gas temperature with radius, from a central value of $1.9 \times 10^8$ K to a temperature of $4.5 \times 10^7$ K at a radius of $1 h_{50}^{-1}$ Mpc, ought to have been observed. The observations show a flat profile (Stewart & Edge 1993). In fact, the temperature *declines* near the centre, presumably due to cooling. We note that a temperature gradient that declines towards the centre will exacerbate the discrepancy between the X-ray and the lensing observations.

Next, we analyze the cluster A1689. A1689 is also an Abell richness class 4 cluster at a redshift of $z_l = 0.17$, with an X-ray luminosity in the 2–10 keV band of $L_x = 2.8 \times 10^{45}$ erg s$^{-1}$. The radial velocity dispersion of the cluster galaxies is approximately 1800 km s$^{-1}$ (Gudehus 1989) and the X-ray temperature of the intracluster gas is about $T_x = 9$ keV or $1 \times 10^8$ K (David *et al.* 1993; Arnaud 1993). As in A2218, the galaxy distribution consists of two seemingly distinct galaxy aggregates. There is a dense concentration where various galaxies appear to have experienced mergers, and smaller concentration approximately 1' away in the NE direction.

The larger of the two clumps is centred on galaxy 82 [we use the labels assigned by Gudehus & Hegyi (1991) in order to identify the galaxies in the cluster]. We designate this galaxy as the centre of the cluster. It is possible that a ROSAT image of the cluster would have helped ascertain whether galaxy 82 is indeed located at the centre of cluster's potential well, but we did not have access to the image. The location of the X-ray peak in the Einstein IPC image of A1689 is consistent the location of galaxy 82; however, it should be mentioned that the resolution of the Einstein image is not very good.

Deep optical images of the cluster, acquired by Tyson, Valdes & Wenk (1990), reveals a few very long arcs of very low surface brightness amongst a large number of weakly but coherently distorted images of faint, blue, presumably background, galaxies. The long arcs are clearly visible when the scaled R-band image of the cluster is subtracted from the B-band image (*c.f.* Figure 2 of Tyson *et al.* 1990). Three of the arcs are located close to the cluster centre and two are located near galaxies 145 and 151 in the secondary clump. Although none of the arcs have a measured redshift, their length, tangential elongation and curvature is very strongly suggestive of the gravitational lensing interpretation.

The two arcs located in the secondary galaxy clump are unlikely to be good indicators of the critical radius of the cluster since they are at quite a distance from the cluster centre (60" and 78") and appear to be affected by the secondary clump. We shall not consider them any further. Of the arcs near the cluster centre, an approximately 10" long arc lies 8" NW of galaxy 62 and at a distance of 47" from the cluster centre. Another arc, a brighter but shorter one, is located close to galaxy 25 at a distance of 43" from the centre. Finally, there is the longest arc in A1689 that occurs near galaxy 167. It is curved around the centre of the cluster, at a radius of 46", and appears to be made up of two arcs with a combined length of 20". We shall use this arc in order to estimate the critical radius. Most likely, it is a fold arc intersecting the critical line. We shall adopt a conservative value of $b = 45''$ for the critical radius. As for the redshift of the background source, we can place an upper limit of $z_s = 4$ since the arc does not appear to have suffered a drop in the B-band flux that would occur when the Lyman break redshifts past $4500\text{Å}$. In fact, we adopt a source redshift of $z_s = 3$. The critical surface density is almost the same for a source located in the redshift range $3 \leq z_s \leq 4$, and a source redshift higher than $z_s = 3$ appears to be implausible because all of the arcs with measured redshifts lie at $z < 3$. If anything, the source is probably at a lower redshift, in which case our estimate of the projected mass required for lensing will be an underestimate. We note that although the shorter brighter arc is only 43" from the centre, we believe that the smaller radius is due to the source being located a lower redshift than the source for the longest arc. This conjecture is supported by the higher surface brightness of the shorter arc. We also point out that the second galaxy clump, if capable of perturbing the lensing configuration of the primary clump, will decrease rather than increase the distance between the longest arc and cluster centre, causing us to underestimate the lensing mass constraint. This effect arises because the arc position angle is roughly perpendicular to that of the second clump and therefore, the arc is on the minor axis of the quadrupole moment generated by the secondary subcluster.

In Figure 2, we show the Einstein IPC X-ray surface brightness profile of A1689. Alongside, we plot the X-ray profile for an isothermal gas of temperature $T = 2.25 \times 10^8$ K in a cluster potential defined by Model 1 mass distribution ($\gamma = 1.51$, $r_c = 37''$, $F_{cD} = 0$). As in Figure 1, the solid line is the surface brightness profile for the model, and the crosses show the result of convolving the model with an Einstein IPC point spread function (approximated by a gaussian with FWHM of 98"), integrating the flux in radial bins, and adding noise. The fit to the observations is acceptable; however, the temperature of the gas is a factor of 2.25 higher than the observed value. The observed surface brightness profile is far too flat and extended to be compatible with the distribution of gas in pressure equilibrium at the observed temperature in a cluster potential where the amount of projected mass within the radius of the arc is given by equation (2). At this point, it is important to note that if the arc is actually at a redshift lower than $z_s = 3$, then the discrepancy in the gas temperature between the model and the observed value would be larger than the factor of 2.25. Even apart from the uncertainties introduced by the lack of information regarding the redshift of the arc, the current best-fit value for the slope of the mass distribution at large radii ($\gamma = 1.51$) is unrealistically shallow. In order to obtain a more realistic mass profile, the gas temperature needs to be higher than a factor 2.25 times the observed value. For example in order to obtain a good fit to the X-ray surface density profile with a mass profile characterized by $\gamma = 2$, the gas temperature needs to be a factor of 2.7 greater than that observed. Similar problems arise even when we assume that the mass profile is given by Model 2 (not shown).

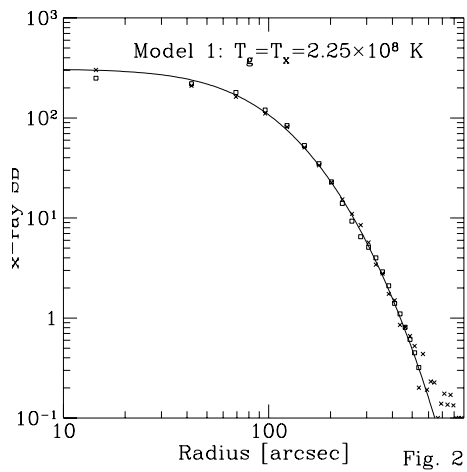

**Figure 2** — Same as Figure 1 but for A1689. We use Model 1 cluster mass distribution and assume an isothermal gas of temperature $T = 2.25 \times 10^8$ K. Again, a good fit can only be obtained when the gas temperature is set at $2 - 2.5$ times larger than the observed value.

The last cluster that we examine is A2163. This cluster differs markedly from A2218 and A1689 in terms of both its X-ray and optical properties. A2163 lies at a redshift of $z_l = 0.2$ and has an X-ray luminosity in the 2– 10 keV band of $L_x = 7 \times 10^{45}$ erg s$^{-1}$; it is one of the most luminous clusters known. It is also the hottest cluster known, with an X-ray temperature of $T_x = 13.9$ keV or $1.6 \times 10^8$ K (David *et al.* 1993). The Einstein IPC image of A2163 is also unusual in the sense that it is not symmetric. (We did not have access to a ROSAT image of this cluster either.) Furthermore, its radially-averaged X-ray surface brightness profile is very flat at small radii and shallow at large radii. In contrast to its X-ray properties, the optical properties of A2163 are quite unassuming. The cluster galaxies is not very compact and the central galaxy is not a bright cD. A2163 is classified as an Abell richness class 2 cluster. Deep optical images of the cluster reveal two arcs around the central galaxy, at a distance of $15''\!.6$ (Tyson 1993, priv. communication). We adopt a value of $b = 15''$ for the critical radius of the cluster. The sources that have been distorted into the arcs lie at $z = 0.728$ and $z = 0.742$ (Soucail 1993, priv. communication); for the purposes at hand, we adopt a source redshift of $z_s = 0.73$. We note that because of the small critical radius and the high X-ray temperature of the gas in the cluster, we expect that there should be no problem in constructing a viable model of the cluster that reproduces the observed surface brightness profile.

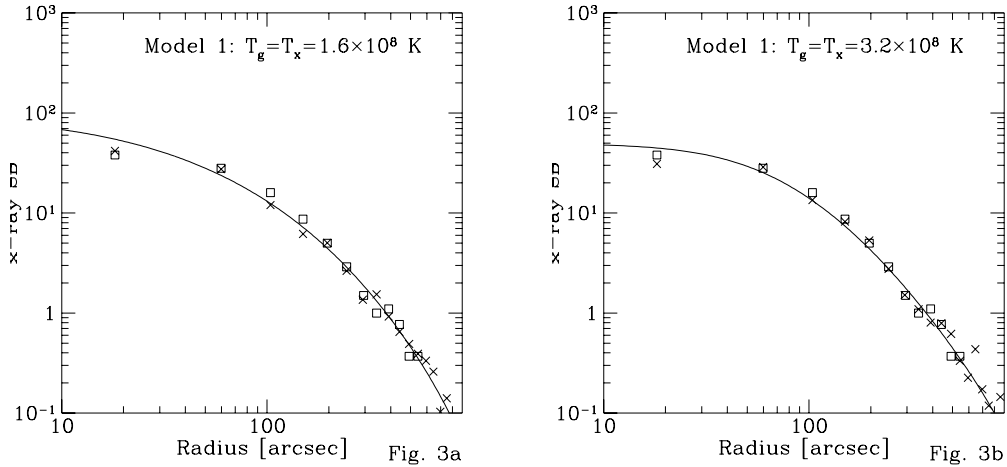

**Figure 3** — Same as Figure 1 but for A2163. In this case, a tolerable fit to the observations can be obtained using isothermal gas at the observed temperature, $T = 1.6 \times 10^8$ K, and Model 1 mass distribution (Figure 3a). When the temperature is increased by a factor of two, the fit improves.

In Figure 3a, we plot the Einstein IPC X-ray surface brightness profile of A2163 alongside the X-ray profile of an isothermal gas at the observed temperature ($T = 1.6 \times 10^8$ K) within a cluster potential defined by Model 1 mass distribution characterized by $\gamma = 1.64$, $r_c = 0$ and $F_{cD} = 0$. As described previously, the solid line traces the surface brightness profile for the model, and the crosses show the result of processing it using our approximation to the Einstein IPC point spread function. As expected, the model results reproduce the observations reasonably well even though the gas temperature is fixed at its observed value. There are some small discrepancies, however. The observed profile is more extended, exhibiting an excess, with respect to the model results, at approximately $100''$. In Figure 3b, we show the improvements resulting from increasing the gas temperature by a factor of 2 above the observed value. The model surface brightness profile matches the observations; the fit is better than in Figure 3a, in that the deviations are consistent with noise. Once again, we have assumed a Model 1 mass distribution, with the best-fit profile being characterized by $\gamma = 1.9$, $r_c = 38''$. In spite of the fact that the latter model offers a better fit to the observed surface brightness profile of A2163, we cannot exclude the case where the gas temperature is the same as that observed, as we could in the previous two cases.

## 4. DISCUSSION

In the generally favoured bottom-up scenarios where structure in the universe forms hierarchically by gravitational instability, clusters of galaxies form via merging of smaller mass concentrations. Simulations of clusters, which include both dark as well as baryonic matter, show that as hierarchical clustering proceeds from subcluster to cluster scale, the associated gas suffers repeated shock heating (*c.f.* Evrard *et al.* 1993, Katz & White 1993, and references therein). Each new merger disturbs the state of the gas and it would seem natural to expect that some turbulent and bulk motions must be present. These state-of-the-art hydrodynamic simulations suggest that in general such motions are not important and that the gas is mostly supported against gravitational collapse by thermal pressure. It is, however, worth noting that simulation results are very strongly dependent on the range of physical processes incorporated, and the resolution of the simulations.

The determination of cluster mass based on X-ray observations rests on the assumption that the gas in real clusters is in hydrostatic equilibrium solely due to thermal pressure. X-ray observations can be used to establish the temperature of the intracluster gas; however, they do not provide any observational tests of the hypothesis that the gas is pressure-supported. With the freedom to choose the underlying total mass density profile, one can fit any combination of observed X-ray temperature and surface brightness profile. The thermal pressure-supported gas hypothesis can only be tested using some independent constraint on the mass distribution. The properties of the galaxy distribution do not provide strong constraints. The uncertainties associated with whether or not galaxies in clusters suffer velocity bias and whether the galaxy orbits are isotropic or not, coupled with the possibility that the optial measurements may be in error as a result of projections effects or be affected by substructure in the clusters, introduce large uncertainties in the cluster mass estimates (Babul & Katz 1993). The gravitationally lensed images, on the other hand, offer a direct means of probing the total mass distribution in clusters, and of testing whether the hot intracluster gas is in thermal pressure-supported hydrostatic equilibrium.

Of the three clusters that we have analyzed, the results for one of the clusters, A2163, are ambiguous. On one hand, the cluster X-ray profile is marginally consistent with both the lensing constraints and the gas temperature being pegged at the observed value. On the other hand, the observed X-ray profile is also consistent with a model where the gas temperature is twice as large as that observed. It should, however, be borne in mind that A2163 is a rather unusual cluster. If as is commonly assumed that the X-ray properties of the cluster offer a measure of the depth of the potential well, then it is somewhat surprising to find only two arcs, both implying a rather small critical radius, in one of the hottest, most luminous clusters known. On the other hand, the lensing results are not unexpected if the galaxy distribution is assumed to reflect the mass distribution in the cluster, in the sense that galaxy/mass distribution is not sufficiently compact. This then raises an interesting question regarding whether the mass distribution in a cluster is best reflected in its X-ray properties or in its optical properties. The only way of answering such a question would be to map out the total mass distribution in clusters using the weakly distorted background galaxies (Kochanek 1990; Miralda 1991; Kaiser & Squires 1993) and compare the resulting map directly with maps of the X-ray and optical light. We note, in passing, that our analyses of the three clusters suggests that the distribution of light in the cluster provides a reasonable picture of the compactness of the total mass distribution.

Coming back to A2218 and A1689, we find that it is not possible for the gas to be supported against gravity solely by thermal pressure. For the gas to be thus supported, it must either have a temperature that is a factor of 2–2.5 times greater than the observed value, if the gas is isothermal, or there must exist a temperature gradient that ought to be detectable. We consider below various other explanations for resolving the discrepancy:

*(i)* The large mass estimate obtained from the giant arcs might be consequence of our having adopted an $\Omega = 1$ cosmological model for universe. However, the inferred mass with a given observed critical radius depends only on the distance ratio $D_{ls}/D_s$, which is very insensitive to cosmology (*e.g.* Blandford & Kochanek 1988). We also note that since the mass estimates derived from the lensing and the X-ray analyses both scale with $H_o$ in the same way, the discrepancy cannot be attributed to our chosed value for the Hubble constant.

*(ii)* The mass estimates based on the lensing argument may be overestimated because the projected cluster potentials may differ significantly from the simple circularly symmetric model that we have adopted. This seems especially likely in the case of A2218, whose central galaxy is highly elliptical. If the ellipticity of the central galaxy reflects the ellipticity of the cluster potential, then the critical line is also elliptical, and the arc that we have used is close to the major axis of the critical line. Using the major axis in equation (2) means that we have overestimated the mass. Even if the cluster potential is not elliptical, the influence, if any, of the secondary clump in the cluster will be to increase the separation between the arc under consideration and the central galaxy. However, according to a detailed model of A2218 that takes into account the presence of two mass clumps and assigns to each the same ellipticity as that of their central galaxies, and that can reproduce the observed arcs in the cluster (Kneib *et al.* 1993b), the total projected mass within the radius defined by our chosen arc is only a factor 1.2 lower the value we have derived. Taking this into account still means that there is a factor of two too much mass for the gas to be supported by thermal pressure. In the case of A1689, the central galaxy concentration is, to a good approximation, spherical and as we have noted previously, the presence of the second clump cannot increase the separation between the cluster centre and the arc that we have used. If anything, it will cause the separation to decrease, implying that our mass estimate is a lower limit. Also, it is likely that the arc lies at a lower redshift than assumed, once again implying that our mass estimate is a lower limit.

(*iii*) The clusters A2218 and A1689 may be highly prolate, with the long axis aligned along the line of sight. From amongst a sample of clusters of some fixed total mass, those that are highly prolate and well-aligned along the line of sight have a higher probability of producing arcs. For example, consider a spherical cluster and stretch it along the line of sight by a factor $q > 1$ while compressing it along the other two perpendicular axes by a factor $q^{-1/2}$. The total mass as a function of radius, therefore, remains unchanged. However, the stretching and the compression results in an increase in the surface mass density, in comparison to the spherical case. The actual amount by which the surface mass density is enhanced depends on the mass density profile in the cluster, but it will always be less than the factor $q$ — the stretching leads to an increase of factor $q$ while the decrease due to compression depends on the mass profile. The probability of producing arcs depends only on the projected surface density. Furthermore, highly prolate clusters that are aligned along the line of sight will appear richer and more compact in optical as a result of the increased galaxian surface density. This may explain why compactness of the galaxy distribution in the clusters is a good indicator of whether the cluster will contain highly distorted images of background sources. On the other hand, the distribution of gas in prolate clusters is also prolate and although the ellipticity would not be observable in the X-ray images due to the cluster alignment, the radial surface brightness profile ought to appear more concentrated towards the centre than that for a typical cluster. This does not appear to be the case. In any case, the resolution of the discrepancy between the X-ray and the lensing mass estimates requires that the axis ratio of the mass distribution be as large $\sim 4$, with the major axis very well aligned along the line of sight. This is a rather rare combination of configuration and orientation and it is rather unlikely that prolateness and alignment is the correct explanation for the discrepancy, unless A2218 and A1689 happen to be rare examples of clusters where such discrepancies arise.

Alternatively, the projection effect may be due to superposition of two mass clumps along the line of sight. If the two clumps are of nearly equal mass, and the gas in each subcluster is in hydrostatic equilibrium, undisturbed by the other clump, the temperature of the gas would only reflect the mass of a single subcluster, while both mass clumps would contribute towards gravitational lensing. Although a double-clump structure is observed in both A2218 and A1689, it is unlikely that the two clumps are acting in unison in order to produce the lensing. The angular separation of the subclusters in each of the clusters places them outside each other's critical radius and also, the central mass concentration in A2218 and A1689 appears to be much more massive than the secondary concentration, as deduced both from the light observed in each clump and from the modelling of the various observed arcs. Therefore, the observed subclusters do not appear to be sufficiently well aligned to explain the discrepancy. On the other hand, it is possible that an unobserved mass clump may be involved; after all, there is mounting evidence suggesting that the present-day clusters are conglomerates of subclusters that are presumably in the process of merging. For example, the ROSAT HRI image of cluster A370 has revealed two previously unresolved peaks that are coincident with the two central galaxies (Böhringer and Mellier 1993, priv. communication), confirming the expectations regarding the nature of the cluster's potential based on the fact that the two central galaxies are both equally luminous, equally extended and that the lensing models of A370 requires a double-peaked surface mass distribution (Kneib *et al.* 1993a). Also, many of the clusters in the ESO cluster redshift survey (*c.f.* Mazure *et al.* 1991) show features that imply a great deal of substructure (P. Katgert 1993, priv. communications).

*(iv)* Substructure in the clusters may also help solve the mystery of the discrepant mass estimates if mass clumps manage to impart bulk motions to the gas in the central regions as they fall through. Hydrodynamical simulations of clusters suggest that during quiescent periods, the gas at large radii is primarily supported by thermal pressure; at small radii, however, the situation is less certain (Tsai *et al.* 1993). Similarly, the response of the gas to subclumps ploughing through the main cluster body has not been well-studied. If bulk motions are present, they can help support the gas, allowing the gas to form an extended core without being hotter than observed. The main problem with this explanation is that unless the bulk motions are coherent (e.g., rotation), they will quickly dissipate. This explanation would, therefore, imply that the presence of large X-ray cores is a transient feature caused by bulk motions imparted to the gas by recent or ongoing mergers of substructure in the clusters.

*(v)* The gas may be supported in part by magnetic pressure in equipartition. Determinations of the general cluster fields using Faraday rotation measures of background or cluster radio sources suggest that overall the intracluster magnetic field is too weak to have any important dynamical influence (Sarazin 1992). In the central regions, however, magnetic fields could be enormously amplified by compression and inflow associated with cooling flows. If magnetic fields are present, they would have to varying on very small scales in order to satisfy the limits on Faraday depolarization (Kim *et al.* 1991).

*(vi)* The intracluster medium in the central regions of the cluster may be inhomogeneous. A multiphase medium has advantage of allowing for a high effective temperature (i.e. pressure divided by average gas density) that could account for the extended gas core, while not being in conflict with the lower temperatures derived from the emission-weighted X-ray spectra (Schwarz *et al.* 1992). A multiphase medium with high temperatures, however, is constrained by the upper limits on the excess of high energy photons in X-ray spectra, with respect to the emission from a single temperature gas (Allen *et al.* 1992).

## 5. SUMMARY AND CONCLUSIONS

In this paper, we examined three clusters in which long arcs have been observed and whose X-ray properties have been well studied. We find that the mass estimate derived from gravitational lensing considerations can be as much as a factor of 2.5 larger than the mass estimate derived an analysis of the X-ray observations. We have considered various possible explanations for the discrepancy. Of these, those that are most promising, and therefore worthy of further consideration, fall into one of the following three categories:

(1) Projection Effects, where either the clusters themselves are highly prolate and extremely well-aligned along the line of sight, or the cluster consists of at least two main clumps that are aligned sufficiently well so as to be hidden by projection in the X-ray maps.

(2) Inhomogeneous Intracluster Medium, where the gas, particularly in the central region, is a multiphase medium.

(3) Non-thermal Pressure Support, where the gas, particularly in the central region, is supported in part by bulk motions and/or by magnetic fields.

If the resolution is the result of either (2) or (3) above, then a substantial modification of our assumptions regarding the physical state, the dynamics, the evolution of the intercluster medium (and the clusters in general) is implied.

In order to unravel further the mystery underlying the discrepancy, the sample of clusters for which both lensing and X-ray observations (temperature and surface brightness profile) are available, needs to be enlarged and thoroughly analyzed. The following observations would be most relevant:

(1) High-resolution X-ray images of the central regions of the clusters — in order to test for the presence of substructure in the X-ray emissivity.

(2) Acquisition of more arc redshifts — in order to allow for improvements to existing lensing models.

(3) Measurements of cluster temperatures — in order to improve mass estimates derived from X-ray observations. This may prove difficult for the faint high redshift clusters, but the X-ray luminosity - temperature relationship may be used.

(4) Measurements of the high energy spectrum of a large number of clusters — in order to test for the presence of a hot phase in the intracluster gas and thereby, constrain the multiphase medium hypothesis.

(5) Observations of the weakly distorted background galaxies — in order to reconstruct the surface density of the clusters (see Kaiser & Squires 1993), and improve the cluster mass distribution models based only on constraints provided by the long arcs. Since weak lensing analysis provides a mass estimate with an approximately constant relative accuracy with radius (Miralda 1991), it should be possible to check whether the discrepancy between the lensing and the X-ray mass estimates extends to large radii. If so, it is possible that part of the explanation for why the apparent fraction of baryonic mass in clusters, as derived from the X-ray data, is much larger than the cosmological baryon fraction implied by the standard theory of cosmic nucleosynthesis in an $\Omega = 1$ universe (see Babul & Katz 1993, White *et al.* 1993 and references therein) may be due to the fact that the X-ray mass estimates are actually underestimates.

On the theoretical side, the hydrodynamical simulations of clusters need to be improved to the point where the treatment of the gas in the central regions of the clusters is realistic and reasonably well resolved. Such simulations are needed to determine the influence of substructure and the importance of bulk motions. If the X-ray surface brightness profile of the gas in the simulated clusters have smaller cores than observed, this may indicate that not all of the important physical processes in the intracluster medium have been taken into consideration.

**Acknowledgements:** We would like to thank Tony Tyson for allowing us access to unpublished lensing data for A2163, and Genevieve Soucail for informing us on the redshift measurements. We are also indebted to Alastair Edge and Gordon Stewart for providing us with the ROSAT image of A2218 in advance of publication. In addition, Alastair Edge helped us obtain the Einstein data for A1689 and A2218, and was a source of useful comments. Finally, we would like to acknowledge many stimulating discussion with Hans Böhringer, Carlos Frenk, "Raja" Guhathakurta, Yannick Mellier, Martin Rees, Peter Tribble, and Simon White. A.B. acknowledges the hospitality of the Institute of Astronomy (University of Cambridge) over the course of his visits during the summers of 1992 and 1993. JM thanks SERC for support in Cambridge, and the W. M. Keck Foundation for support in Princeton.

TABLE 1: CLUSTER PROPERTIES

| Cluster | z | $L_x$(2–10 keV) ($\times 10^{45}$ erg s$^{-1}$) | $T_x$ (keV) | R | $\sigma_r$ km s$^{-1}$ |
|---|---|---|---|---|---|
| A1689 | 0.17 | 2.8 | 9. | 4 | 1800 |
| A2163 | 0.20 | 7.0 | 13.9 | 2 | |
| A2218 | 0.17 | 1.0 | 6.7 | 4 | 1370 |

Notes to Table 1:
X-ray data from David *et al.* 1993
Radial velocity dispersion: A1689 (Tyson *et al.* 1990); A2218 (LeBorgne *et al.* 1992)